\documentclass[final]{raa06}           
\usepackage{graphicx,times}             
\usepackage{natbib}
\usepackage{amssymb,amsmath}
\bibpunct{(}{)}{;}{a}{}{,}
\usepackage[colorlinks=true, citecolor=blue]{hyperref}%

\usepackage{graphicx,kantlipsum,setspace}
\usepackage{caption}
\usepackage{graphics,epsf}
\usepackage{amsmath}                
\usepackage{amsfonts}               
\usepackage{amssymb}                
\usepackage{epsfig}                 
\usepackage{appendix}
\usepackage{graphicx}
\usepackage{float}
\usepackage{color}
\usepackage{multirow}
\usepackage{colortbl}
\usepackage[para,online,flushleft]{threeparttable}
\usepackage{xcolor}

\hypersetup{citecolor=blue, 
            linkcolor=red, 
            menucolor=blue, 
            urlcolor=blue}  

\newcommand{\cm}{{~\rm cm}}

\newcommand{\km}{{~\rm km}}
\newcommand{\s}{{~\rm s}}

\newcommand{\g}{{~\rm g}}

\newcommand{\erg}{{~\rm erg}}
\newcommand{\yr}{{~\rm yr}}

\newcommand{\AU}{{~\rm AU}}

\begin{document}

   \title{On the nature of jets from a main sequence companion at the onset of common envelope evolution
}

   \volnopage{Vol.0 (20xx) No.0, 000--000}      
   \setcounter{page}{1}          

   \author{Noam Soker
    }

   \institute{Department of Physics, Technion, Haifa, 3200003, Israel;   {\it   soker@physics.technion.ac.il}\\
\vs\no
   {\small Received~~20xx month day; accepted~~20xx~~month day}}

\abstract{
I consider a flow structure by which main sequence companions that enter a common envelope evolution (CEE) with giant stars might launch jets even when the accreted gas has a sub-Keplerian specific angular momentum. I first show that after a main sequence star enters the envelope of a giant star the specific angular momentum of the accreted gas is sub-Keplerian but still sufficiently large for the accreted gas to avoid two conical-like openings along the two opposite polar directions.  I suggest that the high-pressure zone that the accreted gas builds around the main sequence equatorial plane accelerates outflows along these polar opening. Most of the inflowing gas is deflected to the polar outflows, i.e., two oppositely-directed jets. The actual mass that the main sequence star accretes is only a small fraction, $\approx 0.1$, of the inflowing gas. However, the gravitational energy that this gas releases powers the inflow-outflow streaming of gas and adds energy to the common envelope ejection. This flow structure might take place during a grazing envelope evolution if it occurs, during the early CEE, and possibly in some post-CEE cases.  This study increases the parameter space for main sequence stars to launch jets. Such jets might shape some morphological features in planetary nebulae, add energy to mass removal in CEE, and power some intermediate luminosity optical transients. 
}

 \authorrunning{N. Soker}            
\titlerunning{Jets from a main sequence companion in CEE}  
   
      \maketitle

\section{Introduction} 
\label{sec:intro}

There are many bipolar and elliptical planetary nebulae (PNe; including some post-asymptotic giant branch nebulae) with central binary systems (e.g., \citealt{Miszalskietal2019MNRAS487, Oroszetal2019, Jones2020}). 
The morphologies of many of these nebulae suggest that a main sequence companion accretes mass from the asymptotic giant branch (AGB) progenitor of the PN and launches jets that shape the PN (e.g., \citealt{Morris1987, Soker1990AJ, SahaiTrauger1998, RechyGarciaetal2017, AkashiSoker2018, Balicketal2019, Derlopaetal2019, EstrellaTrujilloetal2019, Tafoyaetal2019, Balicketal2020, RechyGarciaetal2020}; for an alternative view for some bipolar PNe see, e.g., \citealt{Baanetal2021}). 
In some cases the progenitor of the PN is  a red giant branch (RGB) star (e.g., \citealt{Hillwigetal2017, Jonesetal2020, Jonesetal2022, Jonesetal2023} and \citealt{Sahaietal2017} for the Boomerang nebula). 
The companion can launch the jets before the common envelope evolution (CEE), during, or after the CEE (e.g.,  \citealt{Tocknelletal2014, Guerreroetal2020, Soker2020Galax, Kimeswengeretal2021}).
\cite{BlackmanLucchini2014} study the kinetic properties of outflows from 19 pre-PNe (taken from \citealt{Bujarrabaletal2001} and \citealt{Sahaietal2008}). They conclude that main sequence companions to the AGB progenitors of the pre-PNe should accrete at a very high rate to explain these kinetic properties. Accretion might in some pre-PNe be via a Roche lobe overflow (RLOF), or, more likely for most or all pre-PNe, during a CEE. 

Theoretical studies should reveal the processes that allow main sequence companions to PN progenitors to launch energetic jets, before, during and/or after the CEE. This is the goal of this study. 

Based on the arguments listed above for PNe, as well as other systems, like intermediate-luminosity optical transients (ILOTs) and supernova impostors, I argued in  \cite{Soker2020Galax} that main sequence stars in binary systems can accrete mass at a high rate from an accretion disk and launch jets, and that such high mass accretion rates can take place in the CEE phase, in particular at the onset of the CEE, including a possible grazing envelope evolution (GEE) phase. 
In this paper I examine a possible flow structure that allows a main sequence star to accrete mass and launch jets. 

Several CEE studies consider jets from a neutron star (NS) companion (e.g., \citealt{ArmitageLivio2000, Chevalier2012, MorenoMendezetal2017, LopezCamaraetal2019, Schreieretal2019inclined, LopezCamaraetal2020MN, Hilleletal2022FB, Soker2022Rev, Schreieretal2023}). Other consider jets that a main sequence companion in a CEE launches (e.g., \citealt{ShiberSoker2018, Shiberetal2019, LopezCamaraetal2022, Zouetal2022}).  
It is much more difficult to envision the launching of jets by a main sequence companion in a CEE than by a NS or a black hole (BH) because the accretion flow structure onto a main sequence star in a CEE differs from that onto NS/BH in two significant manners. (1) NSs and BHs have much smaller radii than a main sequence star and therefore a centrifugally supported disk is formed around a NS/BH in a CEE according to the above studies (but see \citealt{MurguiaBerthieretal2017} for difficulties). This is not the case for a main sequence star accreting mass during a CEE because of its much larger radius. (2) The accreted mass onto a NS/BH in a CEE loses energy in neutrinos, and the energy of the accreted mass onto a BH can in addition be accreted by the BH. The accretion process onto a main sequence star in a CEE has no energy sinks. 

\cite{Shiberetal2016} study the removal of accretion energy from accretion disks by magnetic fields. They argue that this might allow a high mass accretion rate onto main sequence stars, up to $\approx 10^{-2} M_\odot \yr^{-1}$ 
for solar type stars, and up to $\approx 1 M_\odot \yr^{-1}$ for very massive main sequence stars. 
In the present study I consider the case of sub-Keplerian accretion flow that forms an accretion belt rather than a centrifugally-supported accretion disk (section \ref{sec:DiskToBelt}) and where the energy removal is by the jets even when the role of magnetic fields is small (section \ref{sec:AccretionFlow}).  
The flow structure that I study might facilitate the launching of jets by a main sequence star in a CEE, at least while the main sequence companion spirals-in in the outer zones of the giant envelope where the flow is not in a steady state (section \ref{sec:Timescales}). 

{ I base  my study in part on earlier papers that study some of the processes that I also discuss here. In  \cite{Soker2004AM} I discussed the role of jets that a main sequence companion launches in a CEE in facilitating envelope removal. That paper does not study the inflow-outflow structure in a sub-Keplerian accretion process as I study here. \cite{SchreierSoker2016} studied the launching of jets by a sub-Keplerian accretion belt as a result of magnetic field amplification. They did not specifically present the flow structure as I do here. The present work is complementary in a sense to their study as here I study an outflow that is powered by thermal pressure rather than by magnetic fields. Both effects of thermal pressure and magnetic fields are likely to operate together in reality. \cite{JiaoWu2011} do study in detail the inflow-outflow structure in a sub-Keplerian accretion flow. I present a similar flow structure here. However, I differ from them in presenting the flow parameters in a way that fits CEE. For example, I include a study of the relevant timescales for launching of jets in CEE (section \ref{sec:Timescales}). } 
I put all these processes together in section \ref{sec:Summary} where I discuss the implications of such an accretion process that launches jets.

\section{Accretion disk to accretion belt transition}
\label{sec:DiskToBelt}

I consider a secondary main sequence star of mass $M_2$ that enters the envelope of a giant star, which might be an RGB star, an AGB star, or a red super giant (RSG). The secondary star accretes mass from the envelope in a Bondi-Hoyle-Lyttleton (BHL) type accretion flow. 
In \cite{Soker2004AM} I derived a crude condition for the accreted mass to form an accretion disk around the secondary star. The assumptions there, as I adopt here as well, are as follows. 
(1) The accretion process is a BHL type accretion flow. (2) The relative velocity between the secondary and the envelope is $v_r = [G M_1(a)/a]^{1/2}$, where $a$ is the orbital separation and $M_1(a)$ is the giant mass inside radius $r=a$. Namely, $v_r$ is approximately the Keplerian velocity of the secondary inside the envelope. (3) The three-dimensional hydrodynamical simulations of \cite{Livioetal1986} that were performed in a rectangle box hold for the CEE, despite that they  did not include the circular orbital motion of the secondary star and not the gravity of the giant. For a density gradient perpendicular to the relative velocity between the secondary star and the envelope in the form $\rho=\rho_0(1+y/H)$ \cite{Livioetal1986} find that the specific angular momentum of the accreted gas is { (see also equation 5 in \citealt{Soker2004AM}) }
\begin{equation}
j_{\rm acc}=\frac{\eta}{4 H} 
\frac{ (2 G M_2)^2}{v^3_r}, 
\label{eq:jacc}
\end{equation}
where $\eta \simeq 0.25$. 
{ Here the density gradient is along the radial direction from the center of the giant star $r$, which is perpendicular to the orbital motion of the secondary star in the common envelope. } 
For an envelope profile of 
\begin{equation}
\rho_{\rm env} (r) \propto r^{-\beta}
\label{eq:EnvDensty}
\end{equation}
at $r=a$ the density gradient has $H=a/\beta$. 

The condition for the formation of an accretion disk (i.e., supported by the centrifugal force alone) is $j_{\rm acc} >j_2$, where $j_2$ is the specific angular momentum of a Keplerian motion on the companion’s equator. Another way to express this condition is to refer to the radius $R_{\rm d2}$, which is the radius of the disk that the accreted gas would form with its average specific angular momentum (the subscript `2' indicates that the accretion disk is around the secondary star). The condition for the formation of an accretion disk { in a CEE, i.e., $a<R_1$ where $R_1$ is the giant's radius, } is then
\begin{equation}
R_{\rm d2} = \frac {j^2_{\rm acc}}{GM_2}=\eta^2 \beta^2
\left[ \frac{M_2}{M_1(a)} \right]^{3} 
a > R_2 ,
\label{eq:EtaBeta1}
\end{equation}
where $R_2$ is the radius of the secondary star.  { Equation (\ref{eq:EtaBeta1}) is as equation 7 in \cite{Soker2004AM}, which was scaled for $\beta=2$. This equation serves as a starting point to analytical studies of the accreted angular momentum in CEE. In the rest of the paper I differ from \cite{Soker2004AM} in discussing the flow structure and in concentrating on sub-Keplerian specific angular momentum.   }

Consider a main sequence star that enters the envelope of a red giant (AGB/RGB/RSG). In the very outer envelope the density profile is very steep, i.e., $\beta \gg 1$, becoming $\beta \simeq 2$ in the main part of the envelope. In the outer region of the envelope, inner to the steep density profile, there is even a density inversion. 
To enter a CEE the main sequence companion cannot be too massive. If it is, then it brings the giant rotation to a stable synchronisation with the orbital motion. { Approximately, to enter a CEE the system should be unstable to the Darwin instability, i.e., $\mu<3I_{\rm env}/a^2$, where $\mu=M_1M_2/(M_1+M_2)$. For $\beta=2$ the envelope moment of inertia is $I_{\rm env}=(2/9)M_{\rm env}R^2_1$.  
Taking a grazing orbit, i.e., $a=R_1$, the condition to enter a CEE by the Darwin instability reads  $\mu< (2/3)M_{\rm env}$, where $M_{\rm env}$ is the envelope mass. For the scaling we use here of $M_2=0.25M_1$ the approximate condition reads $M_2 \la 0.8 M_{\rm env}$ for entering a CEE. The situation might be complicate if the companion launches jets that remove envelope mass as the system approaches a CEE. In that case the system might perform a GEE. }

As well, I do not consider a low mass secondary star that has a small influence on the envelope. { The lower boundary is not well defined as it depends on the degree of influence that we are looking for. As I estimate later, a secondary mass of $M_2 = 0.3 M_\odot$ orbiting inside an AGB star of a solar type star liberates an accretion energy with a power of $\dot E_{\rm acc} \simeq 10^5 L_\odot$. The power goes as $\approx M^2_2$. Therefore, for the accretion power to be larger than the luminosity of the giant, $L_1 \approx 10^4 L_\odot$, the secondary mass should be $M_2 \ga 0.1 M_\odot$. For a solar type mass this gives that the companion can even be a brown dwarf. Somewhat lower mass companions are also possible. }

Overall, I consider the range of  $0.1 M_1 \la M2 \la 0.8 M_{\rm env} \approx 0.5 M_1$. Scaling then equation (\ref{eq:EtaBeta1}) gives 
\begin{equation}
\frac{R_{\rm d2}}{R_2} = 0.39 \left( \frac{\eta}{0.25} \right)^2 
\left( \frac{\beta}{2} \right)^2
\left[ \frac{M_2}{0.25 M_1(a)} \right]^{3} 
\left( \frac{a}{100 R_2} \right) .
\label{eq:EtaBeta2}
\end{equation}

When the main sequence secondary star is outside the envelope and accretes via a RLOF type accretion or when it just enters the outskirts of the giant envelope, i.e., $a$ is in the range $\simeq (0.9- 1) R_1$, an accretion disk is formed around  the secondary star because $\beta \gg 1$. As the companion enters deeper into the envelope, but still in the outer parts, say $a \simeq 0.5-0.9 R_1$, then as $a$ decreases so does $\beta$ and the scaling of equation (\ref{eq:EtaBeta2}) shows that the accretion disk forms very close to the secondary surface and then the accreted gas becomes sub-Keplerian. Namely, centrifugal forces alone cannot support the gas from reaching directly the secondary star and a thermal pressure builds around the accreting secondary star. 

Because centrifugal forces alone cannot hold anymore against gravity, the accreted gas spreads to the two sides of the equatorial plane and forms an \textit{accretion belt} rather than an accretion disk. Namely, centrifugal forces and pressure gradients together support the accreted gas around the secondary star. I consider that such accretion belts can launch energetic jets (e.g.,  \citealt{SchreierSoker2016}). 

The sub-Keplerian accretion will start for higher values of $\beta$ than what the scaling in equations (\ref{eq:EtaBeta1}) and (\ref{eq:EtaBeta2}) give. The reason is that the negative jet feedback mechanism in CEE implies that the mass accretion rate is below the BHL accretion rate (e.g., \citealt{Soker2016Rev, Gricheneretal2021,  Hilleletal2022FB, LopezCamaraetal2022}). A smaller accretion rate implies a smaller ratio of the BHL accretion radius to the density scale-height $H$, and therefore, by equation (\ref{eq:jacc}), a smaller value of $j_{\rm acc}$.    
Indeed, \cite{LopezCamaraetal2022} find in their simulations of CEE of a main sequence companion inside a RGB stellar envelope that the specific angular momentum of the accreted mass is not large enough to form an accretion disk. In section \ref{sec:AccretionFlow} I consider a way by which even a sub-Keplerian accretion flow in a CEE might launch jets. 

\section{The sub-Keplerian accretion flow}
\label{sec:AccretionFlow}

The flow structure is more complicated than what I have described in section \ref{sec:DiskToBelt} because the flow itself depends on the ability of the flow to lose energy. \cite{MacLeodRamirezRuiz2015b} conduct a detailed study of the BHL accretion flow by 3D simulations and present thorough discussions with references to earlier studies. They find that when the adiabatic index of the flow is $\gamma=1.1$ an accretion disk is much more likely to form than for $\gamma=5/3$. The $\gamma=1.1$ mimics an accretion flow that can lose energy. Here I attribute this energy loss (valve as \citealt{Chamandyetal2018} term it) to the jets that carry the energy. However, the simulations of \cite{MacLeodRamirezRuiz2015b} take the central accreting body to be a sink, i.e., they impose a spherical absorbing boundary condition surrounding the central point mass. This is different from a main sequence star that does not have an absorbing surface. Therefore, I do not expect simulations with absorbing inner boundary conditions to have the flow structure as I present here. I encourage CEE simulations with a main sequence star as the accreting mass.  

\cite{MurguiaBerthieretal2017} find in their wind tunnel simulations (rather than full CEE simulations) that accretion disks form only for lower effective $\gamma$, namely, when the gas is more compressible. They mention the partial ionization zones of a common envelope as one where the gas is more compressible. The present study is relevant to these outer envelope zones. Their simulations, however, lack two important ingredients that I have here, namely, a non-absorbing accreting body and a full CEE simulation (as they simulate a wind tunnel). The mass accretion rates that simulations find demonstrate the importance of including full CEE simulation. The wind tunnel formalism that some studies employ leads to mass accretion rates that are much below the values according to the BHL accretion flow (e.g., \citealt{MacLeodRamirezRuiz2015a, MacLeodRamirezRuiz2015b, Deetal2020}). On the other hand, full simulations of accretion from a wind, i.e., simulations that include the orbital motion in a binary system, derive the BHL accretion rate \citep{Kashietal2022}. 

{ With these considerations in mind, I turn to study the flow structure. } 

There are many studies of sub-Keplerian accretion flows (e.g., \citealt{JiaoWu2011, GarainKim2023}). 
When the accreting body is a main sequence star (rather than a NS or a BH) and with the conditions that I studied in section \ref{sec:DiskToBelt}, such a flow does not form a thin accretion disk but rather a rotating flow on the surface of the accreting body which is termed an accretion belt, or a slim disk if it extends to larger distances from the accreting body. These two papers (\citealt{JiaoWu2011, GarainKim2023}) as examples, study the inflow of a sub-Keplerian flow and present the outflow along the two opposite polar directions. The flow structure I present here is qualitatively similar to their study.  
I also note that some observations support the notion that sub-Keplerian disk can launch jets (e.g., \citealt{Nandietal2018, Banerjeeetal2020, Debnathetal2020, Debnathetal2021}). 

In what follows I use Greek letters to mark coordinates around the center of the secondary star, $\Delta$ will be the distance from the center of the secondary star and $\varpi$ the distance from the symmetry (polar) axis with respect to the center of the secondary star. I keep $R_{d2}$ to be the radius of the disk measured from the center of the secondary star even when $R_{\rm d2}<R_2$, and $R_2$ to be the radius of the secondary star. 

\begin{figure}[t]
	\centering
\includegraphics[trim=10.0cm 10.5cm 0.0cm 0.50cm ,clip, scale=0.76]{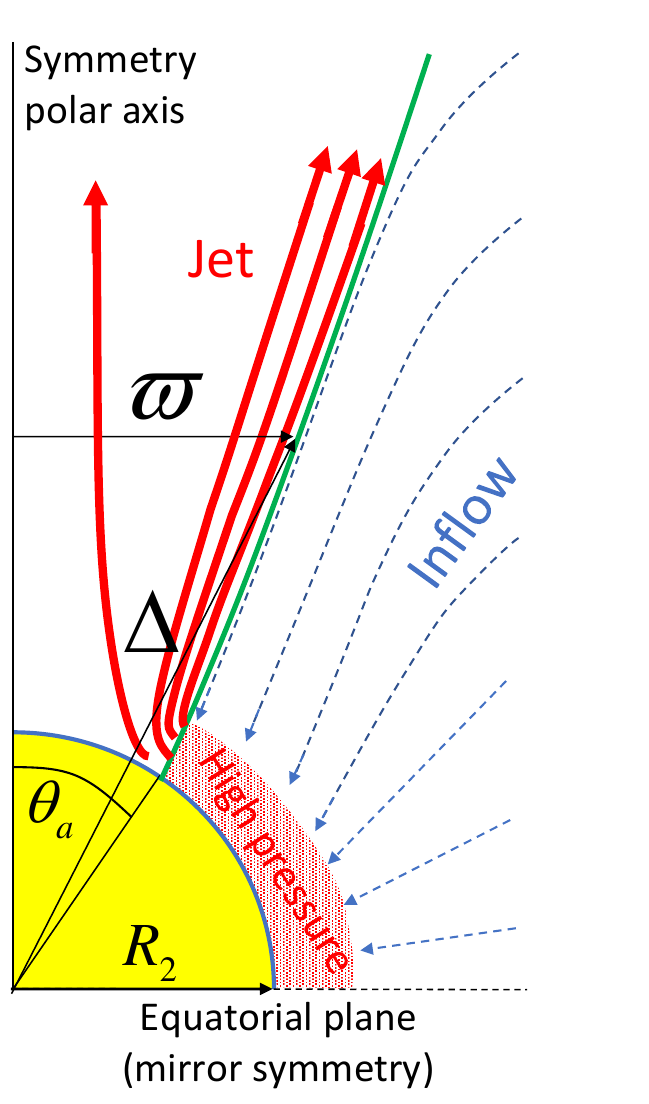}
 \\ 
\caption{A schematic drawing of the sub-Keplerian accretion flow onto the secondary star (yellow region) with the suggested jets drawn in one quarter of the meridional plane. There is an axial-symmetry around the polar (vertical) axis and a mirror symmetry about the equatorial (horizontal) plane. Magnetic fields also play roles in launching the jets but they are not drawn. The dashed-blue lines depict the sub-Keplerian inflow while the solid-red lines depict the outflow, i.e., the formation of jets (only one jet is drawn). The green line is the boundary of the inflowing gas according to equation (\ref{eq:boundary}) for $R_{\rm d2}=0.1 R_2$.  }
\label{Fig:Schematic}
\end{figure}

The situation is as I present in Fig. 
\ref{Fig:Schematic}. Magnetic fields do play a role in launching the polar outflow, i.e., jets (e.g., \citealt{SchreierSoker2016, Shiberetal2016}) but I do not show them in the figure. Note that I present a laminar flow structure. However, simulations (e.g., \citealt{GarainKim2023} for sub-Keplerian accretion onto a black hole) show that the flow can be turbulent. { Most accretion disks, including those that launch jets, are turbulent. The main effect of the turbulence in regards to the jets is the amplification of magnetic fields (acting together with the differential rotation). As magnetic fields can support jets' launching, the presence of turbulence will add to the powering of the jets.  } 

I consider the case where the radius of the secondary star is much smaller than the BHL accretion radius, i.e., $R_2 \ll R_{\rm BHL} \simeq 2GM_2/v^2_r$. { A typical ratio is }
\begin{equation}
\frac{R_{\rm BHL}}{R_2} \simeq 100
\left( \frac{R_2}{0.5 R_\odot} \right)^{-1} 
\left[ \frac{M_2}{0.25 M_1(a)} \right]
\left(  \frac{a}{100 R_\odot} \right).
\label{eq:RadiiRatio}
\end{equation}

I therefore consider that most of the material is flowing onto the secondary star with the average specific angular momentum $j_{\rm acc}$ that corresponds to a Keplerian disk at radius $R_{\rm d2} < R_2$. Namely, a centrifugally-supported disk cannot form. Consider then a parcel of gas with a specific angular momentum $j_{\rm acc}$ that is accreted above (or below) the equatorial plane and it is at $\Delta>R_2$. There is a minimum distance $\varpi(\Delta,j_{\rm acc})$ below which the parcel of gas cannot flow towards the symmetry axis with respect to the secondary star (vertical black line in Fig. \ref{Fig:Schematic}). This distance is given by the balance of the gravitational component perpendicular to the equatorial plane, $G M_2 \varpi/\Delta^3$ with the centrifugal force per unit mass $j^2_{\rm acc}/\varpi^3$. The distance $\varpi$, neglecting forces due to pressure gradients perpendicular to the symmetry axis (perpendicular to the angular momentum direction), is then  
\begin{equation}
\varpi(D,j_{\rm acc})= \left( \frac{R_{\rm d2}}{\Delta} \right)^{1/4} \Delta ,
\label{eq:boundary}
\end{equation}
where $R_{\rm d2}$ is given by equations (\ref{eq:EtaBeta1}) and (\ref{eq:EtaBeta2}). 
On the surface of the secondary star there is an avoidance cap with an angle $\theta_a$ from the symmetry axis that is given by  
\begin{equation}
\sin \theta_a = \frac {\varpi(R_2,j_{\rm acc})}{R_2}= 
\left( \frac{R_{\rm d2}}{R_2} \right)^{1/4} . \label{eq:Alpha2}
\end{equation}
In Fig. \ref{Fig:Schematic} I draw this angle and $\varpi(\Delta, j_{\rm acc})$ for $R_{\rm d2}=0.1R_2$. 

I note here that \cite{JiaoWu2011} consider self-similar flow in the radial direction, use full hydrodynamical set of equations, and solve for the inflow/outflow structure. Because of their self-similar solution the angle that separates between inflow and outflow is constant with distance from the mass-accreting body, i.e., they have $\theta _a (\Delta)={\rm constant}$. They present the variation of this angle with several flow parameters, but not with the specific angular momentum of the accreted gas as I use here with the goal of understanding CEE. For these two reasons it is hard to compare the two solutions. The solution of \cite{JiaoWu2011} gives the flow properties in most of the space, but not close to the symmetry axis. However, it is not simple to apply their solution to CEE. 

Equations (\ref{eq:boundary}) and  (\ref{eq:Alpha2}) teach us that there are wide openings (two opposite openings) along the polar directions that allow the gas to flow out. These openings are not completely devoid of gas because a fraction of the accreted gas does have a low specific angular momentum (as mixing of the inflowing gas is not complete). However, the density there is low and I assume that the accelerated outflow continues to flow out.  

The high-pressure region (red-dotted area in Fig. \ref{Fig:Schematic}) accelerates gas towards the low-density zones near the polar directions. This reduces the pressure around the mass-accreting secondary star and acts as a \textit{valve} to allow the high inflow  rate process to continue. { If the central star is an absorbing body (which it is not here), then the inflow rate is the accretion rate and it can even } be super-Eddington (e.g., \citealt{Chamandyetal2018}).{ However, with a main sequence accretor in a CEE most of the gas flowing into the high pressure zone is deflected to form the outflow rather than being accreted. Only a small fraction of the gas in the high pressure region is actually accreted (see below).  }   

I also note that magnetic fields play roles in accelerating jets by accretion belts (e.g., \citealt{SchreierSoker2016}). The magnetic fields channel rotational energy and turbulence to kinetic energy of the outflowing jets. In the flow structure I study here that takes place inside a common envelope the thermal pressure around the mass-accreting secondary star might dominate the role of magnetic fields. 

In the flow structure that I study here there is no energy loss beside the outflow (jets). There is no neutrino cooling as in the accretion process onto a NS, there is no energy sink as in the accretion process onto a BH, and the entire region is optically thick and there is no energy loss in radiation as in active galactic nuclei. Therefore, the energy is either stored in the accreting main sequence star or is carried away by the outflow (jets). Simulations of accretion flows without radiative cooling (e.g., \citealt{Stoneetal1999, Mosallanezhadetal2021}) show that most of the energy is carried away by the jets and that most of the mass flowing in is diverted to the polar outflow. Super-Eddington accretion flow also show this property (e.g.,  \citealt{Jiaoetal2015, Jiao2023}). 

The typical outflow velocity of jets is the escape velocity from the accreting object $v_{\rm esc}$. { In the specific flow that I study here it will be somewhat below the escape speed. } This implies that the specific energy of the outflowing gas (the jets) is about equal to the binding energy of the gas near the surface of the accreting body, and is about equal to the specific energy of the accreted gas. Therefore, most of the inflowing gas is ejected back (e.g., \citealt{Stoneetal1999}). Only a small fraction of the inflowing gas is actually accreted by the main sequence star { because the high pressure zone deflects most of the inflowing gas to the polar outflow. }  
{ Consider for example a case where a fraction $0.2$ of the inflowing gas is accreted and a fraction of $0.8$ is ejected back. The mass that is actually accreted onto the main sequence star $M_{\rm acc,MS}$ releases an energy of 
$E_{\rm acc,MS} \simeq (0.5)M_{\rm acc,MS} v^2_{\rm esc}$. If the jets (outflowing gas), which are four times as massive as the actually accreted gas, carry this energy, their terminal velocity (far from the main sequence star) will be $v_{\rm jet} \simeq 0.5 v_{\rm esc}$. For an accreted fraction of $0.1$ the terminal velocity of the jets is $v_{\rm jet} \simeq v_{\rm esc}/3$. These velocities are in the range of $\simeq 200-300 \km \s^{-1}$ for typical main sequence stars.}

The flow is basically a large circularization flow involving a mass streaming rate of about the BHL accretion rate $\simeq \dot M_{\rm BHL}$ which is driven by a small accreted mass $\dot M_{\rm acc} = \epsilon \dot M_{\rm inflow} \simeq \epsilon \dot M_{\rm BHL}$ where $\epsilon \ll 1$. 
The outflow mass rate is then $\dot M_{\rm jets} = (1 - \epsilon) \dot M_{\rm inflow}$. Numerical simulations of accretion flow onto main sequence stars in CEE will eventually determine the value of $\epsilon$. I will scale with a value of $\epsilon \simeq 0.05- 0.1$ based on calculations of super-Eddington accretion flows onto BHs (e.g., \citealt{Jiaoetal2015}).  

Since the outflow mass rate in the jets is about equal to the inflow mass rate, the specific angular momentum of the gas in the jets is about equal to that of the inflowing gas, $j_{\rm jet} \simeq j_{\rm acc}$. This implies, by equation (\ref{eq:boundary}), that most of the material in the jets flows close to the boundary along $\varpi(\Delta,j_{\rm acc})$ as is shown in Fig. \ref{Fig:Schematic} by the dense three red lines and as calculations of sub-Keplerian accretion (e.g., \citealt{JiaoWu2011}) and also simulations of super-Eddington accretion flow (e.g., \citealt{AsahinaOhsuga2022}) show.  

\section{Timescales}
\label{sec:Timescales}

{ I now calculate and discuss three different timescales to  further reveal the inflow/accretion/outflow properties of the studied flow. }

During the rapid plunge-phase of the CEE the accretion flow is not in a steady state because the spiralling-in timescale $a/\dot a$ is on about the dynamical timescale of the giant star (e.g., \citealt{Ohlmannetal2016a, Iaconietal2017b, LawSmithetal2020, GlanzPerets2021a, GlanzPerets2021b, GonzalezBolivar2022, Lauetal2022a, Lauetal2022b, Morenoetal2022, Ondratscheketal2022, Tranietal2022, RoepkeDeMarco2023} for some recent papers and references to earlier studies therein). I take the plunge-in  timescale as $\tau_1 = 2 \pi R^{3/2}_1 (GM_1)^{-1/2}$, i.e., the orbital Keplerian time of a test particle on the giant surface. I take therefore $\dot a = R_1/\tau_{\rm Kep}$ where $R_1$ is the unperturbed giant radius at which the secondary starts the plunge-in phase. 
I take again the relative velocity between the secondary star and the envelope to be the local 
Keplerian orbital velocity $v_r$. The BHL mass accretion rate is then 
\begin{equation}
\dot M_{\rm acc}=4 \pi \sqrt{G M_2}  \rho_0 R^\beta_0
\left[ \frac{M_2}{M_1(a)} \right]^{3/2} 
a^{3/2-\beta} ,
\label{eq:Macc}
\end{equation}
where $R_0$ is the radius, measured from the center of the primary giant star, at which the density is $\rho_0$ (the location $R_0$ is not important). 

Since the mass of the primary $M_1(a)$ decreases as the secondary spirals-in the accretion rate increases faster than $a^{3/2-\beta}$ as $a$ decreases in the spiralling-in process. This implies that the ram pressure of the inflowing gas that hits the high pressure zone near the secondary star increases with time. This is true at a given distance from the secondary, because if the high-pressure zone expands the inflow velocity and density decrease and so is the ram pressure of the inflowing gas.

The second timescale is the one during which the accretion flow establishes itself, $\tau_f(R_{\rm acc}) \simeq R_{\rm acc}/v_r$, where $R_{\rm acc}$ is the accretion radius, i.e., the distance from which mass is accreted. Taking the accretion radius to be the BHL accretion radius and the relative velocity the Keplerian velocity at $a$, gives the ratio of the flow time to the spiralling-in dynamical time
\begin{equation}
\frac{\tau_f(R_{\rm acc})}{\tau_1} \simeq 
\frac{1}{\pi} 
\frac{M_2}{M_1(a)}
\left[ \frac{M_1}{M_1(a)} \right]^{1/2}
\left( \frac{a}{R_1} \right)^{3/2} \ll 1 .   
\label{eq:TaufTau1}
\end{equation}
Because $\tau _f \ll \tau_1$ we can consider a BHL-like accretion flow. However, because the mass inflow rate increases there is no real steady state. Note also that because the feedback mechanism reduces the mass accretion rate, the accretion radius is smaller than the BHL radius that I take here. Therefore, a typical ratio, for $0.1 M_1 \la M2 \la 0.5 M_1$, might be $\tau_f(R_{\rm acc})/\tau_1 \simeq 0.1-0.5$. 

Another relevant timescale is the viscosity timescale of the gas that is circulating on the boundary of the inflow from the polar conical-like zones (green line in Fig. \ref{Fig:Schematic}).   In accretion disks this time scale can be $Q \approx 10-100$ times the orbital period at a given radius in the disk. 
Consider then a circulating flow around the inner boundary of the inflow at distance $\varpi(\Delta, j_{\rm acc})$ from the axis (depicted by the green line in Fig. \ref{Fig:Schematic}). Consider a parcel of gas with a specific angular momentum around the symmetry axis $j_{\rm acc}$, a disk radius $R_{\rm d2} <R_2$, and a distance $\varpi$ given by equation (\ref{eq:boundary}). The parcel of gas at the boundary surface $\varpi$ will complete an orbit on a time of $P(\varpi)=2 \pi \Delta^{3/2}/\sqrt{G M_2}$. I find the ratio of the viscous time $\tau_v = QP(\varpi)$ to the flow time to be 
\begin{equation}
\begin{split}
\frac{\tau_v(\Delta)}{\tau_f(R_{\rm acc})} \approx 
\pi Q & \left[ \frac{M_1(a)}{M_2} \right]^{3/2} 
\left( \frac{\Delta}{a} \right)^{3/2} 
 \\ 
 = 0.25  
 \left( \frac{Q}{10} \right)
& \left[ \frac{M_1(a)}{4M_2} \right]^{3/2} 
\left( \frac{\Delta}{0.01 a} \right)^{3/2} .
\label{eq:TauViscosity}
\end{split}
\end{equation}
The more relevant time to compare the viscosity timescale at $\Delta$ with is the flow time from $\Delta$ to the secondary star, which is shorter than $\tau_f(R_{\rm acc})$ that measures the flow time from the accretion radius to the secondary star. This would give $\tau_v (\Delta) > \tau_f(\Delta)$. 
I conclude that the loss of angular momentum by viscosity does not close the two opposite conical-like openings along the polar directions (the regions inner to the green line in Fig. \ref{Fig:Schematic}). 

I end by estimating the actual mass accretion rate. Typical BHL mass accretion rates in the outer region of AGB stars is $\dot M_{\rm BHL} \approx 0.1  M_\odot \yr^{-1}$ .
For example, the parameters of $M_2\simeq 0.3 M_\odot$, $M_a(a) \simeq 1M_\odot$, $a \simeq 1 \AU$, $\beta \simeq 2$ and an envelope mass of $0.4 M_\odot$ inside a radius of $R=1.2 \AU$ gives a density of $\rho (1 \AU) \simeq 10^{-8} \g \cm^{-3}$. Equation (\ref{eq:Macc}) gives then $\dot M_{\rm BHL} \simeq 0.12  M_\odot \yr^{-1}$. The actual mass that is accreted onto the companion is $\epsilon \approx 0.1$ times this mass. The accretion power of an accretion rate of $\dot M_{\rm acc} \simeq 0.01  M_\odot \yr^{-1}$ onto this main sequence star is $\dot E_{\rm acc} \simeq 6 \times 10^{38} \erg \s^{-1}$, which gives a typical power of $\dot E_{\rm acc} \simeq 10^5 L_\odot$. If the companion accretes and launches the jets for a time period of $\simeq 0.1 - 1 \yr$ in the outer region of the common envelope and the jets' velocity is $\simeq 100 \km \s^{-1}$ (according the power and a mass outflow rate of $\simeq 0.9 \dot M_{\rm BHL}$), then the total momentum in the jets is $p_{\rm 2j} \approx 10^{38} - 10^{39} \g \cm \s^{-1}$. This range is within the observed values of the pre-PNe that \cite{BlackmanLucchini2014} study. 
I conclude that if the flow structure that I study here takes place, it might account for the minor aspects of the shapes of some PNe (but definitely not all).

\section{Summary and Implications}
\label{sec:Summary}

The morphologies of many PNe with central binary stars force us to consider main sequence companions to the PN progenitors that launch energetic jets (Section \ref{sec:intro}). Not during all jet-launching phases do the jets breakout from the envelope to shape the PN (e.g., \citealt{LopezCamaraetal2022}). The present study aims at suggesting a flow structure that might take place during a GEE if it occurs, and during the early CEE, i.e., the early phase of the plunge-in stage of the CEE.
At these phases the specific angular momentum of the accreted gas might be sub-Keplerian with respect to the main sequence secondary star. Namely, the accreted gas cannot form a centrifugally supported accretion disk. The flow structure might be as I schematically draw in Fig. \ref{Fig:Schematic}. 

{ Because the jets can barely breakout from the envelope, I do not expect that the jets that a main sequence companion launches inside the envelope shape large lobes. By large lobes I refer to lobes with a size about equal to, or not much smaller than, the size of the non-spherical structure of the nebula ejected during the CEE. However, if the main sequence star launches the jets in a GEE and during the early CEE phase in the outer envelope, then the jets can form arcs and small lobes in the ejected nebula, i.e., lobes with sizes of $\la 0.2$ times the size of the non-spherical part of the nebula. The shaping of large lobes requires probably the launching of jets by an accretion disk that is formed with larger specific angular momentum, i.e., Keplerian accretion disk. } Future 3D hydrodynamical simulations should reveal the morphological features that different jets form.    

Numerical hydrodynamical, and possibly magnetohydrodynamical, simulations should confirm whether the flow structure that I propose here does indeed take place. It is impossible yet to resolve both the entire CEE volume and the accretion flow around a main sequence companion. Instead, the simulations should impose appropriate boundary conditions and set for an inflow of sub-Keplerian gas at the boundaries of the grid. Such simulations should also conserve angular momentum and have a high resolution to resolve turbulence that is likely to develop. Both the differential rotation and turbulence will form a dynamo that amplifies magnetic fields. For that, the following stage will bee the conduction of high-resolution magnetohydrodynamical simulations. 

Although PNe motivate me to introduce this sub-Keplerian flow structure to launch jets I note the following. Firstly, the jets that shape PNe might be launched only before the CEE, like in a GEE phase, or after the CEE (e.g., \citealt{Soker2020Galax}) because during the CEE the jets are deflected and even choked within the envelope (e.g., \citealt{LopezCamaraetal2022}).  
Nonetheless, the sub-Keplerian accretion that accelerates jets might be in some cases relevant to the pre-CEE evolution, in particular the GEE phase if it occurs. The sub-Keplerian accretion might also be relevant to the post-CEE phase when the main sequence companion accretes mass from a circumbinary disk (e.g.,  \citealt{Soker2020Galax}). Secondly, even if the jets are choked within the common envelope they deposit their energy into the envelope and by that can contribute to the removal of the common envelope, at least when the companion is in the outer parts of the envelope. i.e., the early plunge-in phase. 

As I noted in section \ref{sec:DiskToBelt} the negative jet feedback mechanism in CEE reduces the mass accretion rate to be below and even much below the BHL accretion rate (e.g., \citealt{Soker2016Rev, Gricheneretal2021,  Hilleletal2022FB, LopezCamaraetal2022}). This reduces also the specific angular momentum of the accreted gas $j_{\rm acc}$. On the other hand the jets remove mainly gas from the polar directions. This gas has a lower specific angular momentum to begin with. The removal of gas from the polar directions increases somewhat $j_{\rm acc}$ with respect to an accretion without polar mass removal. 
Overall, I expect that with the launching of jets the specific angular momentum of the accreted gas be smaller than what equation (\ref{eq:jacc}) gives. Namely, the effective value of $\eta$ is smaller than $0.25$. This in turn implies that the condition for a sub-Keplerian accretion flow, $R_{\rm d2}<R_2$, is met for steeper density gradients, namely higher values of $\beta$, i.e., even for up to $\beta \simeq 10$ (see equation \ref{eq:EtaBeta2}), which is the case in the very outer parts of the giant's envelope. 

Most relevant to the present study are the simulations that \cite{LopezCamaraetal2022} conducted of a main sequence star inside the envelope of a RGB star. They find that during a GEE and early in the CEE (the early plunge-in phase) jets might breakout. However, jets cannot break out from the envelope when the companion is deep inside the envelope. Note that they inject the jets manually and do not follow the formation of the jets by the accretion flow. In my study here I suggest a way for such main sequence companions in their simulations to launch jets, as \cite{LopezCamaraetal2022} manually inserted in the simulations, even when the accretion flow is sub-Keplerian

The sub-Keplerian accretion onto a main sequence star that powers jets might be also relevant to some ILOTs if they are powered by jets (e.g., \citealt{SokerKashi2012, Soker2021RAA}).\footnote{I follow \cite{KashiSoker2016RAA} in using ILOTs (for usage of the term ILOT see also \citealt{Bergeretal2009} and \citealt{MuthukrishnaetalM2019}) to include all subgroups that are powered by gravitational energy. These subgroups of ILOTs include intermediate-luminosity (red) transients, red novae, luminous red novae (e.g. \citealt{Jencsonetal2019, Blagorodnovaetal2021}), LBV giant eruptions and SN Impostors. Another general term is gap transients (e.g., \citealt{Kasliwal2011, PastorelloFraser2019}).}
The accretion belt/disk is formed by either a mass transfer onto a main sequence star or by tidal disruption of one star onto the other main sequence star (e.g., \citealt{KashiSoker2016RAA}). It is possible that in some cases the accreted mass in ILOTs is sub-Keplerian. Since jets are very efficient in powering ILOTs (e.g., \citealt{Soker2020ApJ}) the present study increases the parameter space of binary interaction to power ILOTs with jets. 

Overall, the present study suggests a flow structure that increases the parameter space for main sequence stars to launch jets in shaping some planetary nebulae, in adding energy to mass removal in CEE, and in powering some ILOTs. 

\section*{Acknowledgments}
 
I thank Diego López Cámara for very good comments, {{ and an anonymous referee for very detailed comments that substantially improved the manuscript. } This research was supported by a grant from the Israel Science Foundation (769/20) and by the Pazy Research Foundation.



\label{lastpage}

\end{document}